# Fault Lines: Navigating Ethics and Responsible AI Where National Policy Meets Local Practice in Public Sector Transformation

**Sitong Lyu[1,2], Shabnam Taghiyeva[1,2], Mohit Kukadia[1,2], Denis Newman-Griffis[1]**

[1]Centre for Machine Intelligence, University of Sheffield; [2] Blavatnik School of Government, University of Oxford
d.r.newman-griffis@sheffield.ac.uk

## Abstract

The UK government has adopted a pro-AI stance to help transform public service delivery in the face of severe financial pressures, but the path forward to translate this vision into responsible AI practice remains ill-defined. While UK policy is often set at the national level, local government authorities are responsible for most public service delivery, and the rapid advance of AI-first narratives in the public sector is exposing fault lines in knowledge and practice at this national-local interface. This paper examines how responsible AI is interpreted and implemented at the interface between the UK's central government and local authorities, taking the high-stakes area of Special Educational Needs and Disabilities (SEND) as a case study. We present a thematic analysis of 17 semi-structured interviews with policymakers, practitioners, and third-sector professionals, supplemented with documentary analysis, to identify barriers and enabling conditions for responsible AI where national policy meets local practice. We identify five interconnected challenges facing local authorities: shadow usage of AI and data privacy risks, market–government asymmetry in AI provision, insufficient workforce readiness, a lack of standardised definitions and measurements, and gaps in human accountability. For each, participants proposed actionable steps, from strengthening data protection frameworks and rebalancing the market–government relationship to enhancing workforce capacity. Our examination of SEND as a case study brings these challenges into sharper focus, showing how high-stakes decisions affecting vulnerable children and families intensify tensions around accountability, fairness, and human oversight, exposing the limits of a principle-based regulatory approach. We argue that responsible AI in the public sector requires not only national policy adjustments but also structural reforms to institutional capacity, values, and governance mechanisms at the local level.

CCS CONCEPTS • Government technology policy • Socio-technical systems • Computing in government

**Additional Keywords and Phrases:** Public sector AI, Digital transformation, Responsible computing, AI ethics

## Introduction

Since 2023, the UK government has increasingly demonstrated its ambition to become an AI maker rather than an AI taker. From the 2023 AI Regulation Whitepaper (or "A Pro-innovation Approach to AI Regulation") [1] to the 2024 AI Action Plan [2], the national government has repeatedly initiated top-down policy frameworks to ensure the UK can reap the benefits of AI development.

Among all sectors, the UK national government particularly pushes for AI adoption in the public sector, accompanied by significant investments and policy mandates. In 2025, the national government's AI Opportunities Action Plan (later called "the Plan") highlights that "[t]he public sector should rapidly pilot and scale AI products and services and encourage the private sector to do the same" [3]. According to the Plan, AI adoption in the public sector and private sector can reinforce each other, amplifying the positive impact of AI across the entire ecosystem. Specifically, the Plan suggests local governments adopt a "Scan > Pilot > Scale" approach to build AI capabilities, pilot prototype AI products and services, and roll out the successful ones.

However, the key is not merely *accelerated* AI adoption but also *responsible* AI adoption. For responsible AI deployment in the public sector, a persistent gap emerges between national principles and local practice. While UK national policy has emphasized high-level commitments to principles including fairness, transparency, accountability, safety, etc. [4], operationalizing these principles in public sector organizations remains difficult. This disconnect is particularly pronounced in local government: UK local authorities are responsible for delivering day-to-day public services yet are often found to lack commensurate funding or technical capacity to implement responsible AI policies set nationally [5]. As a result, responsible AI deployment in local UK councils is frequently negotiated informally based on public servants' professional judgements rather than embedded through systematic governance [6].

This paper digs into and endeavors to address this dual policy-practice and national-local gap by examining how responsible AI is brought into practice within UK local authorities, and what primary factors have enabled or hindered its adoption. We take as a case study local authorities' exploration on the adoption of AI in Special Educational Needs and Disabilities (SEND) services – a high-stakes policy area characterized by legal obligations, resource constraints, and direct impacts on children and families. By studying the intersection of national ambition and local delivery, we investigate the factors enabling or hindering responsible AI use and the knowledge, resources, and policy actions required to support it.

# Literature Review

Prior literature on the operationalization of Responsible AI (RAI) within the UK public sector encompasses contested definitions [6,7], the UK's distinct non-statutory frameworks [8, 9, 10] and empirical evidence of friction between ambitious rhetoric and on-the-ground challenges [11]. We argue that static ethical pillars often fracture under real-world complexity, necessitating a shift toward adaptive governance that treats responsibility as a continuous, sociotechnical process rather than a compliance checklist.

## Defining and Operationalizing RAI

The proliferation of AI systems, including Generative AI (GenAI), within UK public services necessitates a robust and operational definition of RAI to ensure public trust and mitigate socio-technical harm [12]. Broadly speaking, RAI is defined by the World Economic Forum as "the practice of building and managing AI systems to maximize benefits while minimizing risks to people, society and the environment" [13]. Academics tend to prefer the RAI framing over terms like "Ethical AI" or "Trustworthy AI" in the public sector context because "responsibility" is an institutional trait that can be readily legally extended to governing bodies, reinforcing organizational accountability of government agencies [14]. Though, critics have cautioned that RAI could become performative when "AI ethics offices" lack concrete organizational power to shape product development or decisions [15].

However, while a broad framing of RAI exists, the detailed operationalization of its components remains deeply contested. Fewer than 1% of UK public-sector organizations have fully operationalized RAI, and even among those cases, implementation draws on competing definitions shaped by industry, policy, and academic actors [16, 17]. RAI policies address more than 30 distinct principles, with little consistency across documents [18]. While RAI does not have a unanimously agreed definition in the public sector, the UK's 2023 AI Regulation Whitepaper [1] does list 5 principles underpinning the responsible development and use of AI, including:

**Accountability and Human Oversight.** RAI principles typically include clear accountability mechanisms, especially human actors in the loop [14, 19]. Many recommendations for using AI responsibly emphasise auditability, supported by appeal and redress mechanisms, ensuring that human judgement remains central when consequences for individuals are significant [20, 21, 22]. Some academics argue that effective accountability frameworks should consider both individual-level responsibility and organizational-level culture, emphasizing information sharing and learning from mistakes rather than pure sanctions or punishments [23].

**Fairness and Equity.** Many RAI operationalizations state that human controllers of AI should proactively mitigate intrinsic or historical biases in training data to prevent discrimination [19]. Fairness initiatives must go beyond technical fixes to address deeper structural injustices and the non-neutrality of training data [6]. RAI principles tend to emphasize inclusivity and accessibility, ensuring that technological adoption does not exacerbate existing divisions for vulnerable communities [21, 24].

**Transparency.** Explaining how AI systems function and generate outputs is vital for cultivating public confidence [14]. While essential for trust, transparency mechanisms must balance information quantity with the risk of exposing system vulnerabilities [6, 25]. Practical steps being explored in the UK include piloting mechanisms such as the Algorithmic Transparency Recording Standard (ATRS) to document AI usage across councils [24]. Current research highlights three key areas where robust mechanisms and academic investigation are lacking, leading to a substantial policy-practice gap:

**Operational Capacity and Policy Implementation.** A pervasive weakness is the struggle of resource constrained local authorities to translate complex national guidance into locally relevant, functional governance frameworks [24]. Organizations must develop dynamic capabilities to strategically select and orchestrate assets that ensure continuous adherence to RAI principles [17]. Most LAs currently rely on general project management rather than using or developing dedicated ethical frameworks, signalling a critical deficiency in implementation methodologies and policy integration [24].

**Socio-technical Measurement and Auditing.** While principles are established, there is insufficient consensus or practical tooling for the rigorous, systematic measurement and continuous auditing of RAI principles in real-world services [19]. This includes monitoring AI's socio-technical effects like tracking emergent risks and unintended post-deployment impacts [26]. A significant challenge remains the translation of abstract, high-level principles into actionable tasks for practitioners [7].

**Digital Literacy and Skills.** The successful implementation of RAI is actively threatened by pervasive skills and capacity gaps among LGA staff, encompassing both technical expertise and essential ethical literacy [24]. Future research must focus on scalable, effective strategies for accelerating workforce upskilling and securing funding necessary to overcome these constraints [24].

## AI Policy in National Government

These implementation gaps are exacerbated by a national policy landscape that prioritizes flexibility over prescriptive enforcement, creating distinct governance challenges for public sector bodies.

The UK positions itself as a global leader in RAI governance for the public sector, primarily through its principles-based regulatory approach [27]. Yet, an expanding body of academic and evaluative research complicates this narrative, revealing tensions and implementation challenges that are obscured when considering policy documents alone. The UK's framework emphasizes five non-statutory principles for AI: safety, transparency, fairness, accountability, and contestability, seeking to balance innovation with responsible governance [27]. Policy discourse positions this regulator-led, agile approach as inherently "pro-innovation" and adaptable to sectoral diversity. Academic analyses, however, warn that such voluntarist, sector-led regimes risk creating loopholes if regulators are under-resourced or lack sufficient mandate to enforce ethical standards. Empirical research by Roberts et al. notes that these frameworks may reproduce a deregulatory mindset, which undermines public trust and allows harmful practices due to weak oversight without encouraging innovation [28]. Successful governance requires iterative feedback loops that enable institutional learning and adaptation as technologies evolve [11].

The Ada Lovelace Institute documented persistent transparency and accountability shortcomings in UK public-sector algorithm use [29]. Practical implementation struggles are apparent: systems for algorithmic impact assessment and auditing exist, but actual adoption is sporadic and often lacks rigorous public reporting or redress mechanisms [1]. Empirical work on NHS AI deployment similarly finds frontline staff sceptical and overburdened, citing difficulties of integration, explainability, and accountability across digital systems [30]. Parallel survey evidence underscores that public trust in such institutions is crucial for AI legitimacy, showing that transparency and accountability hinge as much on institutional credibility as on technical design [31]. Thus, although policymakers espouse transparency, academic analyses reveal that operational realities deliver limited explainability, inconsistent stakeholder engagement, and patchy accountability in high-stakes contexts.

Compared to the EU AI Act's binding rules [32], and governance frameworks from NIST [26] and OECD [33], UK policy is said to be more principle-driven and less prescriptive, prioritizing innovation over hard regulatory constraints [28, 34, 35]. Researchers argue that while this approach enables UK regulators to iterate responsively, the result is also fragmented oversight and uneven protection levels relative to international peers.

A crucial limitation in both government frameworks and academic assessment is the lack of robust empirical evaluation of AI impacts in "high-stakes" domains – for example, in areas like Special Educational Needs and Disabilities (SEND). Many policy interventions remain at the guidance or pilot stage, without systematic studies on outcomes, equity impacts, or public trust. The literature converges in calling for more comparative, transparent, and longitudinal evaluations that move beyond principles to measurable societal outcomes [6, 36].

UK RAI policy in the public sector aspires to principled flexibility, but these academic and evaluative studies have critiqued its real-world limits, particularly around accountability, enforcement, and evidentiary support for claims of responsible innovation. This gap between aspiration and implementation underscores the continuing need for empirical research on AI in public sector practice, and for more robust, evidence-based and enforceable standards for algorithmic governance.

## AI Implementation in Local Government

AI is increasingly being integrated into the daily operations of local governments to enhance the efficiency of administrative and operational tasks. For example, some local councils use chatbots and automated call centre systems to streamline public inquiries, as well as tools such as Automated Minute Taking and Microsoft Copilot to organize meeting notes or generate work-related content [37, 38]. However, as existing academic research on local governments' AI applications often lags behind the rapid development of AI technologies, it tends to offer a relatively outdated or narrow view of AI in practice. The Local Government Association's (LGA) "State of the Sector: Artificial Intelligence - 2025 Update" indicates that 95% of UK local councils are now using or exploring Generative AI [5]. Furthermore, the report identifies three additional categories of AI beyond "Generative AI": "Perceptive AI," "Predictive AI," and "Simulation AI," which enable functions like advanced computer vision, sensor-based recognition, and predictive analysis based on existing data patterns. This shift shows a widening AI adoption beyond efficiency enhancement in the UK local governments.

Several studies highlight challenges local councils have encountered while using AI. Yigitcanlar, et al., [37] for example, analyses international experiences from countries such as China, the US, the UK, Australia, and Sweden using a grey literature review methodology, identifying problems including system incompatibility, data security, financial constraint, and adversarial attacks [39]. Similarly, Vogl's earlier report also highlights the challenges of data management, unclear problem definition, and limited supplier understanding of local contexts [38]. However, these studies emphasize technical aspects of AI implementation, and pay limited attention to the social challenges of AI adoption in local governments, including severe public mistrust, great fear of labour replacement, unfair digital divide and data insecurity [24].

When narrowing the focus to specific service areas, such as Special Educational Needs and Disabilities (SEND), rigorous academic studies and even grey literature become even scarcer. Aliu reviews AI-based solutions for special education but does not focus on public sector delivery mechanisms [40]. In contrast, the "Chartered College of Teaching", supported by the UK Department for Education, discusses the potential of AI to alleviate workload and resource constraints in SEND provision [41]. It calls for greater local government support in training teachers to use ICT and AI tools, improving access to reliable digital infrastructure, and adopting a collaborative approach between leaders and educators.

## Methodology

This qualitative study examined how RAI principles are enacted and negotiated within organizational and policy practice in the UK public sector. Data collection took place between June and August 2025 – a period marked by rapid policy developments following publication of the UK Government's AI White Paper [1], the AI Opportunities Action Plan [3], and subsequent guidance, including the AI Playbook for Government [42].

We conducted semi-structured interviews with seventeen participants, including practitioners, policy makers, and technical specialists directly involved in the governance or use of AI across the UK public sector. Initial participants were identified through existing collaborations, professional networks, and targeted outreach, extended by snowball sampling. This expert sample was designed to represent institutional diversity and depth of experience, emphasizing interpretive insight over numerical breadth. Table A1 (see Appendix in supplementary material) summarizes participant roles. Descriptions are generalised to preserve anonymity.

We conducted online interviews of 30-45 minutes. Interviews addressed current uses of AI in public services, perceived opportunities and risks, understandings of responsible or "good" use, and organizational capacity for implementation. These themes reflected the framework outlined in the ethics protocol to ensure consistency across the national/SEND and local-government strands.

All interviews were audio-recorded with informed consent, transcribed verbatim, and anonymized prior to analysis. Interview data was complemented with publicly available policy materials, including national policy frameworks, regulatory guidance, audit reports, and research outputs from independent institutes and think tanks. These documents contextualize participants' perspectives within the broader evolution of responsible-AI policy and governance in the UK public sector.

### Data Analysis

Interview transcripts were analysed using inductive thematic analysis. Initial open coding was conducted to identify recurrent patterns in participants' understandings and experiences of RAI within the public sector. Codes were iteratively grouped into higher-order categories reflecting recurring concerns around accountability, fairness, organizational capability, and governance practice.

Interpretive consistency was supported through regular team discussions, where researchers compared coding decisions, reconciled differing interpretations, and refined analytical categories. Findings were synthesized into cross-cutting themes linking national policy discourse with local implementation practices, aligned across policy, organizational, and ethical levels of analysis. Documentary materials corroborated and enriched these themes for an integrated understanding of how principle-based visions of RAI intersect with everyday administrative realities.

## Results

### Challenges

The challenges of using AI responsibly in the UK public sector operate on two interconnected levels: systematically at the national level, and locally within individual local authorities. According to participants, current national-level solutions do not provide sufficient concrete guidance to resolve the same set of problems at the local level; meanwhile, emerging challenges and solutions surfacing within local councils are still waiting to be reflected and coordinated systematically. This section presents five challenges identified by participants.

#### Shadow Usage and Data Privacy Issues

Multiple participants raised concerns about the increasing shadow, unsafe usage of AI tools like ChatGPT and other Generative AI in the local authorities. This challenge has been recognised at the systemic level. In 2024, the UK national government released "Guidance to civil servants on use of generative AI" [43], urging civil servants to "never put sensitive information or personal data into these tools." This guidance was subsequently superseded by "Generative AI Framework for HMG" [44] in 2024 and "AI Playbook for the UK Government" [42] in 2025, which both list out ten principles and highlight the importance of data protection in governments' usage of AI.

Despite the guidelines and playbooks on the national level, participants of our study pointed out that local council staff are using ChatGPT "secretly," "unsafely," or "unethically." They put sensitive, internal case notes into these public AI models, violating the fundamental data stewardship rules. Participant 13, in particular, noted the necessity to tackle this problem:

> "… the increase is significantly over a thousand users a month with the council traffic to ChatGPT … (P13).

These concerns become even more acute in sensitive areas such as social care, children's services, and education. As Participant 9 put it, councils are "the custodians of the law," bound by "very, very tightly… legislation about things that councils have to do and are not allowed to do," especially when handling the most sensitive categories of data such as children's records. This legal environment means that the public sector's approach to AI must be more cautious than the private sector.

**Market-Government Asymmetry in AI Provision**

Participants frequently raised concerns about AI products for local governments being developed by large technology monopolies with enormous market power. Microsoft accounts for over £1 billion in contract value since 2018 - nearly three times the second-largest supplier Palantir and far exceeding the rest of the market combined [46]. This market concentration leads to two primary problems.

Firstly, it leaves the local authorities almost no alternatives or market competition:

> "...the local authority market is... a very small number of specialist providers... there's not a lot of innovation there... what they want to do is sort of apply their own proprietary AI... and then charge us... to use it." (P12)

Several local authorities' digital leads also mentioned the irresponsiveness of AI service providers when they demand new, sector-specific AI feature updates:

> "At the moment, councils are sort of at the mercy of vendors to an extent. They'll often provide feedback … but it's not necessarily guaranteed that it will be done in a timely manner" (P9).

Thus, although generic AI tools prevailed across the public sector, personalized, contextualized AI products meeting specific demands of the public sector are still difficult for the local councils to find.

Secondly, the fact that many AI tools are owned by large technology corporations rather than the public sector seems to strengthen the ethical concern of shadow usage. Participants frequently questioned the ethical standards of AI vendors explicitly, thus being alerted by unregulated reliance on outcomes of Generative AI like ChatGPT.

> "… ownership of a lot of these techs is from the big tech bros … it doesn't feel like it's in the hands of people who've shown themselves to be particularly responsible …" (P11)

**Insufficient Skills and Workforce Preparation**

Another challenge consistently pointed out by participants is that the UK public sector workforce does not have enough experts for the AI-driven changes. P7 cites the statistics published in the 2025 UK State of digital government review: only 2% of local public sector headcount, compared to 6% of the overall UK public sector, had a background in digital data or technology. [47] This implies a systematic lack of "technology literacy" across the local public sector.

Daily experiences of participants resonate with this systematic phenomenon. According to P11 and P12, despite that 95% of local staff are using AI, formalized AI training is absent, forcing them to learn secretly, independently, and unsafely. Even though general guidelines created at national scale exist, they are too difficult to be understood and operationalized.

> "… the AI responsible use policy … long winded full of IT jargon … not very user friendly… The policy is buried somewhere" (P11).

This lack of training and thus understanding of AI also exacerbates fear and mistrust towards AI tools. Several participants expressed the prevailing fear that AI is a tool that will lead to job loss and undercut public servants who are already poorly paid. Also because of this fear of the unknown, local council staff are less motivated to master foundational AI skills like effective prompting, resulting in a lack of capacity and hidden reliance on AI.

**Lack of Standardised Definitions and Measurements**

Additionally, local council staff express frustration at the lack of standardized measurements for AI-driven productivity gains. Although many participants experience positive impacts from AI, including time released, better work-life balance, and enhanced service quality, generating a solid "return on investment" for introducing AI tools into the public office remains difficult. Unlike savings brought by other deployments like new equipment, additional staff, the savings attributed to AI are typically framed in terms of hours saved rather than cashable budget reductions. This under-demonstrates the returns of AI and overlooks other forms of positive influence. As Participant 13 noted:

> "…return on investment is really hard to prove and that's what we really need to prove because as a public authority we can't be seen to be spending public money for the sake of it." (P13)

This measurement gap cuts both ways. The public sector lacks standardized metrics for either positive or negative impacts of AI, particularly around risk assessment. The result is an absence of rigorous cost-benefit analysis for AI procurement in the public sector: local councils procure and adopt AI tools not on the basis of clear, evidence-based assessment but out of anxiety to keep pace with the AI trend. This pattern of adoption without accountability can itself be irresponsible and runs counter to the principles of RAI that national policy espouses. For example, among participants, there is no shared language on assurance baselines, counterfactuals, error rates, rework, backlog, or timeliness, and we are told that these rarely feature in council discussions. Equally, there is limited consideration of trust and equity signals like complaints, appeals, channel shift, disparity checks, or risk indicators like privacy errors, incidents, near misses, in the measurements of AI adoption.

**Lack of Human Accountability**

Finally, one challenge for the UK public sector is the lack of human responsibility and accountability when AI is involved in decision-makings. Participants observe the tendency public servants to overly rely on and trusting the outputs of AI without having deeper reflection on the outcomes. The second point is particularly highlighted:

"research that's coming out of AI … eating into critical thinking skills and decreasing grey matter." (P11)

This is why participants urged the public sector to have "human-in-the-loop" on using AI-generated outcomes. They argue that specify human accountability is crucial to maintain the responsibility of public sector.

"We don't believe at the moment that AI should be used for any... decision making purposes which affects things like... benefits payments or claims." (P12)

"I don't ever want it to replace and make decisions where humans should be involved … there's that understanding of people's scenarios …" (P13)

This emphasis of "human-in-the-loop" was powerfully reinforced by real-world near-miss cases. A participant (P13) described "a prime example" where a social worker used Co-pilot to summarize an assessment, only to find that AI "called out something completely different." They noted that "if she'd not read that, the right care plan might not have been provided for the end user."

## Areas for Action

### Enhancing Data Protection

Participants repeatedly highlighted the significance of consistent data protection frameworks designed and enforced by national government, including Information Commissioner's Office guidance, Data Protection Impact Assessments (DPIAs), and Equalities Impact Assessments (EQIAs). These were seen as the primary mechanisms for managing shadow and inappropriate usage of AI:

"... DPIA or EQIA take you through some of those things, especially if we think about children and SEND a DPIA would be quite comprehensive …" (P9)

However, participants noted that the capacity to apply these frameworks varies significantly across councils. The UK National Audit Office (NAO) [51] and LGA [52] have frequently noted this uneven digital and legal capability across local councils in their reports. As fulfilling DPIA and EQIA does not equate to meeting the highest ethical standard of AI deployment, this capability imbalance results in significant disparities in how responsibly AI is adopted across local authorities:

"...doing best practice is only for councils with a certain amount of capacity …" (P9)

### Regaining Market-Government Balance

Beyond enhancing AI procurement, participants proposed increasing the role of the public sector in developing its own AI tools. In the UK, the Local Government Association (LGA), the representative body for local authorities, is coordinating collective piloting of AI tools and has already found local case studies of councils "building their own bespoke tools … to tackle specific children's … needs" (P9). According to participants, these self-developed AI tools usually accommodate local demands better than one-size-fits-all AI services provided by large tech vendors.

However, participants also recognized the uneven distribution of digital capacity across councils, making it challenging for some to develop in-house AI products. To address this imbalance, participants raised two solutions. The first was the potential role of nationwide organizations like LGA or the national government to develop a nationally owned "sovereign AI" aligned with public good and public sector demands. The 2025 AI Opportunities Action Plan posited the UK as "an AI maker, not an AI taker," explicitly stating the public sector should play a leading role in piloting and scaling AI products across all economic sectors and digitalizing public service deliveries [3]. Nevertheless, this raises concerns of political trust in the government itself: people fear a sovereign AI can have political biases or authoritarian monitors; and financial constraints meant limited investment in government-developed AI.

"... the pressure would be more on how we can build a better sovereign AI that each country can have … but then you get into political distrust…" (P11)

The second solution was the adoption of a cross-sector ecosystem approach to reduce monopolistic control, achieved through nationwide innovation policy, university-led research and development, and coordinated public-private partnerships.

Participants who were already engaged in such collaborations reported tangible benefits:

"I sit on the local government collaboration innovation pilot subgroup… the authorities that are partnering with the universities are the ones that are moving forward at a greater pace …" (P15)

### Enhancing Workforce Readiness

Addressing the market–government imbalance also requires building internal capacity to make use of new opportunities. Participants proposed several ways to fill in the skills gap. Firstly, they argued that local and national governments must raise attention to the recruitment procedure, hiring people with different familiarity levels with using AI:

"So you've got to have people who've got different skill sets… who are really skilled at using it (AI) to the maximum, and people who are skilled at talking about it" (P7).

Correspondingly, recent research suggests that public-sector organizations, including those in the UK, should prioritize skill-based hiring rather than relying solely on formal qualifications for roles in AI and sustainability [48]. Such an approach may help local councils mitigate direct competition with private technology companies, which are often able to offer substantially higher salaries to candidates with traditional high-profile AI credentials.

Finally, many participants advocated for inter-council dialogues and co-training sessions to prepare the public sector for adopting AI. Participant 10 recommended

LGA's information session around the "AI pre-procurement checklist" to manage risk and procure services suitable for each local council's digital portfolio:

> "... a really pleasing element of sharing that you've got councils who are doing similar things, slightly different, covering discrete areas, and how they can learn from each other is the key to make sure that solutions can be rolled out widely." (P7)

**AI in SEND**

Having identified five key challenges and corresponding action areas for RAI governance across local government, this section uses SEND as a case study to illustrate how these challenges are reflected and, in many cases, intensified in a high-stakes policy area directly affecting children and families. SEND is characterized by legal obligations, resource constraints, and decisions that profoundly shape individual lives, making it a particularly consequential domain in which to examine the entangled systemic and local issues of responsibly using AI in the UK public sector. Our analysis traces how the preceding challenges play out in this sensitive sector, and how the UK's principle-based regulatory framework, which emphasises cross-sectoral norms rather than sector-specific mandates, leaves practitioners to navigate these tensions largely on their own.

These challenges are more significant in SEND for two reasons. Firstly, the UK's pro-innovation, principle-based regulatory approach governs AI largely at the point of use, effectively delegating the burden of assurance and risk mitigation to individual sectors, local authorities, schools, and service providers. Rather than prescribing national mandates, this model relies on sector-led interpretation. However, insufficient sector-specific guidance results in highly inconsistent practice. Participants described a widening disconnect between the UK government's stated ambition to harness AI for economic growth and public-service improvement, and the fragmented, resource-constrained reality of implementing these goals within education and SEND. Several participants noted that ministers have limited "headspace" for AI in SEND due to competing national priorities, leaving the sector vulnerable to strategic neglect. This risk of deprioritization was explicitly articulated by one national policy advisor:

> "While the broader system is being built – across labour, industrial, and innovation policy – we need to make sure education doesn't fall behind" (P1).

Secondly, the education sector exhibits structural barriers that compound this neglect. The system is deeply fragmented: academies, schools, and local authorities operate under different governance arrangements, confining innovation to isolated contexts. As governance consultant P2 summarized:

> "The education system is very fragmented… The answers are there, but they're in pockets of good practice, and not many people know about them." (P2)

Participants also characterized the Department for Education (DfE) as risk-averse and "slow to act" on AI, generating hesitation, uncertainty, and uneven adoption in the sector. This inertia undermined what participants framed as the state's "institutional responsibility": central government articulates expectations for RAI use but falls short in providing the sector-specific guidance, resources, or mandates needed to support implementation on the ground. This context of structural neglect and sectoral fragmentation heightens our five challenges in SEND.

**Shadow Usage and Data Privacy**

Shadow usage of AI and data privacy concerns emerged as the most frequently raised issue among SEND participants. In SEND, where the data involved is highly sensitive and the decisions directly affect vulnerable children and families, the risks are considerably more acute.

The absence of sector-specific AI guidance for education leaves SEND practitioners to navigate AI adoption on their own. Without unified templates or clear rules on acceptable use, staff rely on general-purpose generative tools without oversight or training. With staff often left to "teach themselves or copy colleagues" (P3, P5), this unregulated adoption produces significant variability in technical capability, safety, and awareness of ethical risk.

Shadow usage is particularly prominent in the education sector precisely because of the structural dynamics described above: the national deprioritization of education relative to competing policy areas means that sector-specific guidance remains weak or absent. The fragmented nature of the system means that even where pockets of good practice exist, they do not spread, so SEND practitioners operate in a guidance vacuum.

Even where national-level policies do exist, participants described a persistent gap between systemic-level instructions and local implementation. Existing AI policies were described as "anemic" and "not fit for purpose" (P1). As one participant noted:

> "But honestly, having an AI policy doesn't necessarily change what they do in practice when deploying these tools" (P6).

This lack of oversight was characterized as a "lack of leadership and governance" (P5), leading to uneven application of data-protection norms.

Compounding this, participants described a culture of institutional risk aversion that shapes accountability practice in ways that paradoxically increase risk. This defensive posture can inadvertently increase risk, as institutional inertia and delays in providing safe, official alternatives drive staff toward unregulated experimentation and risky workarounds. Participants widely agreed that the core risk stemmed not from the AI technology itself, but from the human behavior and misuse enabled by a lack of guidance and oversight. In the absence of timely, safe, and officially sanctioned solutions, this caution pushes users toward unregulated experimentation such as practitioners pasting sensitive child data into general-purpose generative tools (P6). One consultant highlighted the systemic implications: "the biggest risk for me is inertia" (P2), a sentiment reinforced by others who warned that "the biggest risk is not doing it" (P6).

**Workforce Readiness and Standardized Measurements**

This fragmentation and the shadow practices it enables are compounded by a deeper structural problem: the practitioners navigating these challenges simply lack the capacity to do so effectively. The capacity gap identified earlier as a general challenge across local councils is particularly acute in the SEND landscape. Participants linked these governance gaps to chronic shortages of time, expertise, and funding, describing schools and local authorities as "resource-starved." Consequently, staff frequently learn about AI "in an ad hoc or sporadic way" (P5), and institutions struggle to keep policies updated due to the pace of change. As one consultant noted:

> "You can write an AI policy this week, and next week a new model drops, and everything shifts again." (P2)

Resource-starved schools and local authorities face difficulty delivering training, maintaining compliance, or building the sociotechnical capability required for RAI use. Participants also questioned the competency of leadership, doubting qualifications school leaders have to lead in an area as "vast and complex as AI" (P2). As one SEND framed it, ethical use depends less on compliance and more on staff having the necessary training to understand "what do I need to be wary of and careful about" (P3).

This lack of capacity also undermines the ability to establish standardised measurements for AI use in SEND. Without the workforce readiness to implement and maintain assessment frameworks, meaningful evaluation of AI's impacts remains out of reach. Despite these challenges, participants emphasized that meaningful accountability cannot rely solely on statutory enforcement or one-off compliance checks. Several interviewees argued that oversight must be ongoing, adaptive, and embedded in organizational culture. As one respondent reflected:

> "We can't just launch a tool and walk away. A year later, we need to ask – has anything changed? Do we need to do it differently?" (P6)

This call for continuous review aligns with broader RAI frameworks that view accountability as distributed and iterative rather than static.

**Human Accountability and Oversight**

The question of human accountability takes on a heightened significance in SEND. While national principles emphasize transparency and accountability, participants described ambiguity over who ultimately bears legal and ethical responsibility in practice, particularly regarding who should ensure safeguards, manage risks, and provide redress when automated systems influence high-stakes decisions. Across interviews, participants highlighted a "liability mismatch" between who is formally responsible for AI oversight and who actually has the authority or capability to act.

This mismatch becomes especially consequential in SEND because of the legal requirement for individualization. While participants agreed that AI could increase accessibility and reduce administrative workload, which offers a "baseline level of quality" regardless of staff fatigue, they feared that this efficiency might come at the cost of the individualization required by law. Concerns centered on the risk that standardization would lead to homogenization, with one practitioner noting:

> "Parents' biggest fear is that the nuance of their child will be lost, that all the plans will start looking the same." (P6)

This warning that assessments could become "reductive, no longer about their child" (P3) echoes broader regulatory debates, such as those in the EU AI Act, which classify AI systems used in education and essential public services as "high-risk" due to their potential impact on fundamental rights.

Beyond homogenization, participants also raised the risk of algorithmic bias: AI systems may reproduce structural inequities embedded in training data. One policy advisor noted that existing models are "very Western-centric and probably lacking in diversity" and must reflect the world we live in (P3). This echoes wider critiques around unrepresentative data.

Despite these risks, several participants expressed cautious optimism, arguing that under appropriate governance, AI could enhance fairness by reducing inconsistency and exposing bias. As one put it:

> "You can't see the bias in a human's head, but you can monitor bias in an AI tool" (P6).

Transparency, they suggested, enables contestability and redress, offering a potentially fairer baseline than opaque human decision-making. Others pointed to AI's pedagogical potential, describing prospects for "endlessly patient teaching" for pupils who struggle in conventional classrooms (P3). This nuanced position illustrates a shift among practitioners and organizational leaders from moral panic toward reflective experimentation, precipitated by the recognition that AI can mitigate unobservable human bias and counter the systemic inertia that participants identified as the greater risk.

Participants also consistently returned to the importance of participatory design to ensure legitimacy and trust. This approach is deeply rooted in the disability rights and justice movements, which popularized the central principle that "Nothing about us without us." As one SEND advisor recalled hearing this slogan from youth advocates (P4), this participatory ethic aligns with the Design Justice framework, emphasising that those most affected by the outcomes should lead design processes. Participants demanded genuine co-production by involving parent-carer forums, students, and professional associations. One consultant, who designed an AI drafting tool, stated:

> "I only want to do it if it's a real co-production, not just a box-ticking exercise where you chat with children to say you've done it." (P6)

## Discussion

This study has revealed three cross-cutting themes:

### From National Framework to Local Guardrails

Our findings consistently demonstrate that national-level actions, while necessary, are not sufficient for RAI deployment in local government. The UK government has produced a succession of frameworks, from the 2023 AI Regulation Whitepaper [1] to the AI Playbook [42], yet participants described a persistent disconnect between these high-level documents and the day-to-day realities of AI use in local councils. LGA's 2025 consultation on AI in government [45] illustrates this variation starkly: only 3% of responding councils completely banned generative AI tools on council devices, while others allowed usage with or without limitations. This patchwork of local approaches persists despite national guidance, confirming what Roberts et al. [28] described as a fundamental tension in the UK's principle-based regulatory model: flexibility at the national level produces fragmentation at the local level.

Overall, RAI deployment in local government is less about introducing technology and more about building usable, practice-focused guardrails: role-based training, simple rules for what data can be used where, clear "safe vs unsafe" examples, and application-specific oversight that reflects local context. These require bottom-up development informed by local service demands, institutional capacity, and the specific risks of each policy area, as the SEND case study vividly illustrates. This aligns with prior research that translating abstract, high-level principles into actionable, low-level tasks for practitioners remains one of the most significant challenges in RAI [7], and the finding that most local authorities currently rely on general project management rather than dedicated ethical frameworks [24]. On workforce readiness, a similar gap emerges while the national government has invested in initiatives such as the AI Skills Hub and digital transformation programmes, participants described local councils struggling to recruit and retain AI talent in competition with the private sector. Bone et al.'s [48] research on skill-based hiring suggests a possible path forward, but participants stressed that without locally embedded training and recruitment strategies, national workforce initiatives will continue to fall short.

### RAI as Lens on Pre-existing Problems

The second theme emerging from our findings is that many of the challenges facing RAI adoption are not AI problems. They are long-standing structural issues like uneven digital capacity, chronic underfunding, risk aversion, etc. that predate AI and are merely exposed and amplified by its arrival. This observation echoes the National Audit Office's [51] assessment of the challenges in implementing digital change and the LGA's [52] capability and capacity review, both of which documented these disparities well before generative AI entered the mainstream.

This reveals a structural tension: while data privacy regulation applies uniformly, the ability to enact RAI use varies significantly based on funding, staff expertise, and internal governance structures. As participants noted, "doing best practice" is something that only councils with sufficient capacity can realistically achieve. These limitations entrench uneven capability and inconsistent data practices, echoing the broader concern that AI adoption has "outpaced governance" across the UK public sector. In SEND, participants linked these governance gaps to shortages of time, expertise, and funding, describing schools and local authorities as "resource-starved." Participants suggested that councils must implement systematic reforms to elevate internal capacities before the national government makes more stringent RAI frameworks legally binding.

Beyond internal capacity, participants also stressed that councils must actively communicate with the public and demonstrate transparency to build trust. Several interviewees noted that public engagement is essential. They also pointed to the challenges of digital inclusion, noting that some groups, such as elderly residents or those unfamiliar with digital technology, may struggle to understand or feel confident about AI. Collectively, these reflections underscore that RAI use is not only an internal governance challenge; it also requires deliberate, ongoing engagement with residents to ensure that AI adoption aligns with public expectations and supports trust in local government. This resonates with the Tony Blair Institute's [31] finding that public trust in AI depends heavily on transparency and perceived institutional competence.

### A Structural Shift in Values and Governance

The third and perhaps most fundamental theme is that RAI in the public sector requires not merely adjustments to existing governance structures but a deeper shift in values, perspectives, and institutional mechanisms. Just as previous technological transformations, from industrialization to the rise of the internet, demanded changes not only in tools but in social relations and organizational structures, the integration of AI into public services calls for a qualitative rather than quantitative modifications to traditional works.

Some participants noted that senior-level public servants should receive training on how AI may reshape local authorities' governance structures and working practices, in order to better prepare their workforce for organizational change. Others suggested that council leaders should prepare for shifts in staff workload, mental health, and cognitive capacity as AI begins to take on more daily, mundane tasks. Interestingly, while most participants agreed that AI would alleviate administrative workloads, they expressed differing views on whether public servants' mental health would improve or deteriorate. While Participant 10 believed that staff would feel more stressed, Participants 13, 14, and 15 reported that AI had helped them improve their work-life balance, enabling them to build confidence, self-esteem, and more meaningful relationships with their families. Both observations are captured by previous academic literature on human resource management in the AI era: Carillo et al. [49] on AI's potential to support work-life balance, and Tarafdar et al. [50] on the technostress that new technologies can introduce. Nevertheless, participant point to specific, practical considerations that should be incorporated into senior public servant training programmes, particularly around managing changes to cognitive load, supporting staff well-being, and fostering organizational environments capable of adapting to AI-enabled work redesign.

These findings suggest that the conversation about RAI in the public sector must move beyond compliance and risk mitigation toward a broader rethinking of what public service delivery looks like in an AI-enabled era. This echoes Ibitoye et al.'s [6] call to shift from static ethical pillars toward adaptive, participatory governance, and Akbarighatar's [17] argument that organizations must develop dynamic capabilities to ensure continuous adherence to RAI principles. Without this deeper structural shift, RAI risks remaining an aspiration layered onto institutions that were not designed for it.

### Limitations and Future Work

This study offers interpretive insight into RAI governance during a formative phase of AI adoption in the UK public sector. While the qualitative design enabled in-depth exploration of practitioner perspectives, the purposive and relatively small sample constrains generalizability. Findings should be interpreted as illustrative rather than representative. Future research could build on this work through longitudinal and comparative studies, mixed-method evaluations of AI implementation outcomes, and participatory approaches involving service users, educators, and technology developers, particularly in sensitive domains such as SEND, where the stakes of getting responsible AI wrong are highest.

## Conclusion

This paper examines how responsible AI is brought into practice within the UK public sector, and what primary factors have enabled or hindered its adoption. Our findings identify five interconnected challenges, including shadow usage and data privacy risks, market-government asymmetry, insufficient workforce readiness, a lack of standardized measurements, and gaps in human accountability. The paper also traces how each is reflected and intensified in the high-stakes domain of SEND. Across these challenges, three broader lessons emerge: national frameworks alone are insufficient without locally usable guardrails; many barriers to responsible AI are long-standing structural issues merely exposed by AI's arrival; and meaningful progress requires not just policy adjustments but a deeper shift in institutional values, governance mechanisms, and leadership.

By situating SEND at the intersection of national ambition and local delivery, this paper illustrates that responsible AI in the public sector is less about introducing technology and more about building the knowledge, capacity, and shared frameworks to guide practice. Translating abstract principles into equitable outcomes for children and families, and for public services more broadly, demands the kinds of resourced, participatory, and context-sensitive governance structures that the UK's principle-based approach has yet to deliver.

## Ethics and Reflexivity

This study was undertaken with research ethics approval from the University of Sheffield, administered by the School of Computer Science (protocol 068867). All participants provided informed consent and had the option to withdraw their data at any stage prior to anonymization. Interviews were audio-recorded via Google Meet, with initial automated transcription followed by manual correction of transcripts.

Reflexively, the authors acknowledge their positionality within intersecting policy, governance, and ethical domains. Lyu's experience in education policy and SEND governance informed attentiveness to institutional and regulatory contexts; Taghiyeva's background in AI safety and digital-governance research contributed an organisational and systems-level perspective; and Kukadia's training in law and policy provided an equity-focused and rights-based analytical lens. Regular team reflections were undertaken to surface assumptions, challenge disciplinary biases, and ensure that the analysis centred participants' perspectives rather than researchers' priors.

## Acknowledgments

This study was funded by the University of Sheffield.